\newcommand{\be}{\begin{eqnarray}}
\newcommand{\ee}{\end{eqnarray}}
\newcommand{\nn}{\nonumber \\}
\renewcommand{\thefootnote}{\fnsymbol{footnote}}
\newcommand{\vecb}[1]{{\bf #1}}
\newcommand{\vecg}[1]{\mbox{\boldmath $#1$}}

\documentclass[12pt]{article}

\usepackage{amsmath,amssymb}
\usepackage{graphicx}
\usepackage{cite}
\usepackage{bm}

\def\theequation{\arabic{section}.\arabic{equation}}
\begin{document}

\vskip 15mm

\begin{center}

{\Large Once more on the Witten index  of  $3d$ supersymmetric YM-CS 
theory.}

\vskip 4ex

A.V. \textsc{Smilga}\,$^{1}$,

\vskip 3ex

$^{1}\,$\textit{SUBATECH, Universit\'e de
Nantes,  4 rue Alfred Kastler, BP 20722, Nantes  44307, France
\footnote{On leave of absence from ITEP, Moscow, Russia.}}
\\
\texttt{smilga@subatech.in2p3.fr}
\end{center}

\vskip 5ex

\begin{flushright}
{\it \ldots It is high time, comrade theorist,
to disclose  your \\ magic 
tricks, especially the  one with  the disappearing\\   magnetic flux.
Readers  are worried about its fate \ldots }
\end{flushright}

\vspace{1mm}

\hfill After {\sl Master and Margarita} by M. Bulgakov

\vspace{3mm}

\begin{abstract}
\noindent  
The problem of counting the vacuum states in the supersymmetric 
$3d$ Yang-Mills-Chern-Simons theory is reconsidered.
We resolve the controversy between its original calculation in \cite{Wit99} at large volumes 
$g^2L \gg 1$
and the calculation based on the evaluation of the effective Lagrangian in the small volume limit,
$g^2L \ll 1$ \cite{ja}. 
We show that  the latter calculation suffers from uncertainties associated with 
 the singularities in the moduli space of classical vacua where the Born-Oppenheimer approximation breaks
down. We also show that these singularities can be accurately treated in the {\it Hamiltonian} 
Born-Oppenheimer method, where one has to match carefully the effective wave functions on the
Abelian valley and the  wave functions of  reduced  non-Abelian QM theory 
near the singularities.  This gives the same result as original Witten's calculation.   
\end{abstract}

\renewcommand{\thefootnote}{\arabic{footnote}}
\setcounter{footnote}0
\setcounter{page}{1}

\section{Introduction}

  $3d$ supersymmetric gauge theories attracted recently a considerable attention in view of newly
discovered dualities between certain  ${\cal N} = 8$ and ${\cal N} = 6$ versions of these theories
and string theories on $AdS_4 \times S^7$ or $AdS_4 \times \mathbb C \mathbb P^3$ backgrounds, 
respectively 
\cite{3duality}.

In this paper, we discuss the simplest ${\cal N} = 1$ version of such theories with nontrivial dynamics -
the   supersymmetric YM-CS theory  with the Lagrangian 

 \be
 \label{LN1}
  {\cal L} \ =\ \frac 1{g^2} {\rm Tr} \left \{ - \frac 12 F_{\mu\nu}^2 +
  i\bar \lambda /\!\!\!\!D \lambda \right \} +
  \kappa {\rm Tr}  \left\{ \epsilon^{\mu\nu\rho}
  \left( A_\mu \partial_\nu A_\rho - \frac {2i}3 A_\mu A_\nu A_\rho \right ) - 
\bar \lambda \lambda \right \} \, .
   \ee
The conventions are:  $\epsilon^{012} = 1, \ D_\mu {\cal O}  = \partial_\mu {\cal O}  - 
i[A_\mu, {\cal O}] $ (such that $A_\mu$ is Hermitian);  
   $\lambda_\alpha$ is a 2-component Majorana $3d$ spinor belonging to the adjoint 
representation of the gauge group.
   We choose
   \be
   \label{gamdef}
   \gamma^0 \ =\ \sigma^2,\ \ \ \gamma^1 = i\sigma^1,\ \ \ \gamma^2 = i\sigma^3 \ .
    \ee
 This is  a $3d$ theory and the gauge coupling constant $g^2$ carries the dimension of mass. The physical 
boson and fermion degrees of freedom in this theory are massive,
  \be
  \label{mass}
  m = \kappa g^2\ .
   \ee
 In three dimensions, the nonzero mass brings about parity breaking.  
The requirement for $e^{iS}$ to be invariant under certain large gauge  
transformations (see e.g. Ref.\cite{Dunne} for a nice review)
 leads to the quantization condition
  \be
    \label{quantkap}
    \kappa =  \frac k {4\pi}  \ .
    \ee
with integer $k$. 

The first question to be asked for any supersymmetric theory is whether supersymmetry is 
spontaneously broken and, if not, what is the number of vacuum states. In most cases 
(and, in particular, in this case) the latter coincides with the Witten index
 \be
\label{Witind}
I \ =\ {\rm Tr} \{ (-1)^F e^{-\beta H} \} \, .
 \ee
This index was evaluated in \cite{Wit99} with the result
 \be
\label{IkN}
I(k,N) \ =\  
[{\rm sgn}(k)]^{N-1} \left( \begin{array}{c} |k|+N/2 -1 \\ N-1 \end{array} \right)\ .
 \ee
for $SU(N)$ gauge group. This is valid for $|k| \geq N/2$. For $|k| < N/2$, the index vanishes
and supersymmetry is broken. In the simplest $SU(2)$ case, the index is just
 \be
\label{Ik2}
I(k,2) \ =\ k \ .
 \ee.

The result (\ref{IkN}) was obtained by the following reasoning. Consider the theory in a {\it large}
spatial volume, $g^2L \gg 1$. Consider then the functional integral for the index (\ref{Witind}) 
and mentally perform a Gaussian integral over fermionic variables. 
This gives an effective bosonic action
that involves the CS term, the Yang-Mills term and other higher-derivative gauge-invariant terms. 
After that, the coefficient of the CS term is renormalized 
\footnote{This is for $k >0$. In the following, $k$ will be assumed to be positive by default though the results 
for negative $k$ will also be mentioned. The gauge 
coupling $g^2$ is also  renormalized in some irrelevant way and new couplings (of 
still less relevant higher derivative terms)
appear. },
  \be
\label{kren}
 k \ \to \ k- \frac N2 \, .
 \ee

At large $\beta$, the integral is saturated by the vacuum states of the 
theory, which depend on 
the low-energy dynamics of the corresponding effective Hamiltonian. 
The latter is determined by the term
with the lowest number of derivatives, i.e. the Chern-Simons term, the effects due to the YM term and
still higher derivative terms being suppressed at small energies. Basically, the spectrum of vacuum states
coincides with the full spectrum in the topological pure CS theory. The latter was determined some time 
ago 
 \begin{itemize}
\item by establishing a relationship between the pure $3d$ CS theories and $2d$ WZNW theories 
\cite{WitCMP} 
 \item  by canonical quantization of the CS theory and direct determination of wave functions annihilated
by the Gauss law constraints \cite{Eli}. To make the paper more self-sufficient, we briefly review
the latter method in  Appendix B. 
 \end{itemize}

Then the index (\ref{IkN}) is determined as the number of states in pure CS theory 
with the shift (\ref{kren}).  For example, in the $SU(2)$ case, the number of CS states is $k+1$, 
which gives (\ref{Ik2}) after the shift. 

 In what follows, we will only consider the case $N=2$. A generalization of the analysis to other
groups involves purely technical complications, which are, however, well understood and
not controversial. We refer the reader to Refs.\cite{Eli,ja} for details.

Speaking of the controversy, it arises when the same problem is considered with a different method.
Following the logics of \cite{Wit82}, we considered the theory in a {\it small} volume, $\xi_{1,2} \in (0,L)$, 
 \ $g^2L \ll 1$, with  periodic boundary conditions. The smallness of the parameter $g^2L$ allows one 
to apply the Born-Oppenheimer ideology and to evaluate the effective Lagrangian depending only on
the relevant for low-energy dynamics {\it slow} variables. The slow bosonic variables represent in this 
case zero Fourier modes of spatial components of vector potential with zero classical energy. 
The latter implies that  the field strength $\sim f^{abc} A^{b(\vecb{0})}_1 A^{c_(\vecb{0})}_2$ is zero 
and $A^{a(\vecb{0})}_{j=1,2}$ belong to the Cartan subalgebra.
For $SU(2)$, there is only one Abelian color component and there are only two
bosonic slow variables $C_j \equiv A_j^{\rm Cartan\ (\vecb{0})}$.  
All other modes are fast and can be integrated over.

Is it important that the slow configuration space is compact, the fields $C_j$ varying within the range
$C_j \in (0, 4\pi/L)$. Indeed, a field outside this range can be brought into it by a {\it large}
(i.e. not continuously deformable  to unity, like $U(\vecg{\xi}) = \exp\{2\pi i \frac {\xi_1}L \sigma_3 \}$)
 gauge transformation.  The effective
theory describes then a motion 
\footnote{For other groups, we have the motion over $T \times T$, $T$ being the maximal torus of the group. }
over $T^2 = S^1 \times S^1$ 
 with a, generally speaking, inhomogeneous magnetic
field. This problem was analysed in \cite{Novikov}. For consistency (more exactly, for the spectrum
to be supersymmetric \cite{flux} ), the flux of the magnetic field should be quantized, 
$$\frac \Phi {2\pi} = q = {\rm integer} \, .$$
The Witten index of this Landau-Dubrovin-Krichever-Novikov theory coincides with $q$.

At the tree level (when the fast modes are not integrated over, but just ignored), the magnetic 
field is homogeneous and the magnetic flux is $q = 2k$. 
The vacuum wave functions can in this case be written explicitly, they
are related 
\footnote{We will explain how this comes about in Sect. 4, see Eqs.(\ref{solFzbar})--(\ref{chim})}.
to theta-functions of level $2k$,
 \be
   \label{Psim}
   \chi_m \sim \ \sum_{n = -\infty}^\infty \exp \left \{ -2\pi k \left(n+y + \frac m{2k}\right)^2- 
2\pi i k x y -4\pi i k x 
\left(n + \frac m{2k} \right)
   \right\} \ ,
    \ee
    where 
 \be
\label{xy}
x = \frac {C_{1}L}{(4\pi)}, \ \ \ \ \ \ \ \ \ \ \ \ \ \ \ \ y = \frac {C_{2}L}{(4\pi)}\ . 
 \ee
 and $m = 0,\ldots,2k-1$.  When $k < 0$,
the vacuum wave functions are fermionic, involving the holomorphic factor $\psi$ (a superpartner of $C_j$). 
Not all of the states (\ref{Psim}) are admissible. The gauge invariance of the states in the original theory
dictates that the effective wave functions should be invariant under Weyl reflections. Indeed, such Weyl reflections
for the effective wave functions can be realized as certain large gauge transformations for the wave functions 
of original theory.  For $SU(2)$, Weyl group involves only one element,
$A_j \to -A_j$.
    There are   $k+1$ Weyl invariant combinations:
 $$ \Psi_0,\ \Psi_k, \  {\rm and} \ \ \Psi_m + \Psi_{2k-m} \ \ (m = 1,\ldots,k-1) \ .$$
 Thus, at the tree level, we obtain the value $k+1$ for the index.

This value is modified when taking into account loops. The loops are irrevant in the middle 
of the dual torus, $C_j \in (0, 4\pi/L)$, where the Born-Oppenheimer approximation works well, 
but there are four ``corners''  where the
approximation breaks down and loop corrections are relevant: 
 \be
\label{corner}
C_j = (0,0); \ \ \ C_j = (2\pi/L, 0); \ \ \ C_j 
= (0, 2\pi/L); \ \ \ C_j = (2\pi/L, 2\pi/L) \, .
 \ee 
 We have shown in Ref.\cite{ja} that the {\it fermion} loop
\footnote{It is sufficient to consider a single loop. 
One can argue that the second and higher loops do not contribute.}
 brings
about an extra effective magnetic field with the flux $-1/2$ in each corner and $\Phi^{\rm extra\ fermion}/(2\pi) = -2$
 all together. 
(For $k <0$, the signs of both the tree-level and loop-induced fluxes are opposite.)  This alone
would renormalize the total flux $2k \to 2k-2$, which would give $(k-1)+1$ Weyl-invariant vacuum states
in agreement with (\ref{Ik2}). However, in the framework of this approach, the {\it gluon} loop seems to be
 equally important.
It gives twice as large extra effective magnetic flux as the fermion one, but with the opposite sign. This would give
 the
total flux $2k - 2 + 4$ = $2(k+1)$ and $k+2$ vacuum states with a blatant contradiction with (\ref{Ik2}) !

 The resolution of the paradox goes along the lines anticipated already in \cite{ja}. 
The extra flux comes from the regions around the singular points (the {\it corners} (\ref{corner}) ) where
the Born-Oppenheimer approximation breaks down. This makes the whole analysis precarious.
The {\it raison d'\^etre} for this paper are two simple remarks:
 \begin{itemize}
\item The effective Lagrangian (as any Lagrangian) can be determined only up to a total time derivative. Normally,
such time derivative does not change anything, but if we add a derivative 
$\sim \frac d{dt} \ln [(C_1 + iC_2)/(C_1 - iC_2)] $, which is {\it singular} in the corner $C_j = 0$, 
this brings about an extra 
delta-functional flux, which may change the index.  The effective action method does not control 
well such contributions, which leaves the result for the index uncertain.
\item The corners can still be treated within the effective Hamiltonian method. To be quite precise, the ambiguity
mentioned above displays itself also there and consists in the freedom to multiply the 
fast ground state wave function by a singular factor 
 \be
\label{singfactor}
\sim [(C_1 + iC_2)/(C_1 - iC_2)]^\alpha \ .
 \ee
 The point is, however, is that this ambiguity can be fixed by imposing proper boundary conditions at the corners.
Indeed, in each such region, one can still apply the Born-Oppenheimer procedure, to single out 
a finite number of slow variables and integrate over all other variables. For example, in the region
 near the corner, $A_j =0$ the slow variables represent constant (not necessarily Abelian) 
Fourrier modes  $A_j^{a = 1,2,3(\vecb{0})}$.  The effective Hamiltonian (it is nothing but the original
Hamiltonian dimensionally reduced to (0+1) dimensions) involves, 
thus, 6 bosonic variables. 6 is greater that 2, but, still, 
this problem turns out to be treatable, if capitalizing on the gauge
invariance requirements. 

Matching the vacuum solutions of this corner effective Hamiltonian to the solution of the Abelian valley
effective Hamiltonian (this is possible to do even though the former are not known exactly)
gives us boundary conditions for the Abelian BO wave function and count the number of vacuum solutions.
 Our final result coincides with (\ref{Ik2}).

 \end{itemize}

The rest of the paper is organised as follows. In the next section, we perform an accurate calculation
of gluon loop contribution in the effective action. (In Ref.\cite{ja}, only the calculation of the fermion loop
was described in details.) We show that, indeed, 
$$ \Phi^{\rm extra \ gluon} \ = -2 \Phi^{\rm extra\ fermion} \ .$$
 We discuss then singular total derivative contributions that are difficult to control.
 
In Sect. 3, we analyse the dimensionally reduced QM Hamiltonian near the corner $A_j = 0$, study what
happens with its vacuum wave function near the Abelian valley, $A^a_j \approx C_j \delta^{a1}$, and calculate the 
effective valley 
Hamiltonian. The latter involves  Pancharatnam-Berry (PB)phase \cite{PB}  --- an extra gauge potential in the space $\{C_j\}$ of slow
variables arising after integrating out the fast ones. We show that, using the most natural definition of what
is understood under the effective wave function, this PB phase is associated only with the fermion factor in the 
fast wave function and  brings about the contribution $-1/2$ to the flux from each corner.

In Sect. 4, we show how, irrespectively of the ambiguity associated with including or not a factor like
(\ref{singfactor}) in $\Psi^{\rm fast}$, the requirement of regularity for the  wave functions at the corners 
allows one to find them on the full dual torus and count them. 
Extra gauge fields dwelling in the corners
modify both the form of the wave functions (they are now given by  Eq.(\ref{Psiwithfluxes}) below) and their counting.
Before imposing the Weyl invariance requirement, we have  $2(k-1)$ rather than $2k$ functions 
of which only $k$ are left
when Weyl invariance is imposed.  

There are four technical Appendices. Appendix A is purely mathematical being devoted to theta-functions.
In Appendix B, we remind how the states were counted in pure CS theory. 
In Appendix C, we accurately study the behavior of the non-Abelian 
ground state wave functions near a corner.
In Appendix D, we construct the corner Hamiltonian with explicitly reduced gauge constraints.

\section{Effective action.}
\setcounter{equation}0

Let us discuss first the renormalization of the theory (\ref{LN1}) in the infinite
volume. It was studied earlier \cite{Rao,Kao} in covariant gauges. One can use, alternatively,
the Hamilton gauge $A_0 = 0$, in which case the gluon propagator is
 \be
\label{Green3d}
D_{jk}^{ab}(\omega, \vecb{p}) \  = \frac {ig^2 \delta^{ab} } {\omega^2 - \vecb{p}^2 - m^2} \left[ \delta_{jk} - 
\frac {p_j p_k}{\omega^2} - \frac {im}\omega \epsilon_{jk} \right]
\ .
 \ee
This choice simplifies the calculations in the gluon sector (no ghosts and only one graph to evaluate). The known
result
 \be
\label{renk}
k \to k - \left( \frac N2 \right)_{\rm ferm. \ loop} + \left(  N \right)_{\rm gluon \ loop}
\ =\ k +  \frac N2
 \ee
is, of course, reproduced. 

To evaluate the effective action in the small finite volume, we note first that the corrections are only large near 
one of the corners. If choosing, say, the region near $A_j = 0$, we notice that one should only take into account
the zero Fourrier modes of the gluon fields propagating in the loop - nonzero modes 
have masses $\sim 1/L$ and their contribution is suppressed for small volumes. Thus, we can neglect the spatial 
dependence of the fields and perform the calculation in the dimensionally reduced theory with the Lagrangian
\be
\label{Lag}
 L = \frac 1{2g_0^2} (\dot A^a_j)^2 + \frac m{2g_0^2} \epsilon_{jk} \dot A_j^a A_k^a 
- \frac 1{4g_0^2} [(A^a_j A^a_j)^2 - A^a_j A^a_k A^b_j A^b_k]    \nonumber \\
- i\frac {\epsilon^{abc} }2 \left[ \bar \psi^a \bar \psi^b A^c_+ +  \psi^a  \psi^b A^c_- \right]
 + m \bar\psi^a \psi^a  \ ,
 \ee
where
\be
\label{gQM}
g^2_0 \equiv  g^2_{1d} = \ \frac{  g^2_{3d}}{L^2} 
\ee 
and $A^a_\pm = A^a_1 \pm i A^a_2$ (and similarly
for other vectors below). 
 We assume the Abelian background 
to be directed along the first color axis, $A_j^1 \equiv C_j$. Then $A^{a=2,3}_j$ are the fluctuations. An
inspection of the quadratic in $A^{a=2,3}_j$ part of the Lagrangian,
 \be
\label{LquadrA}
g^2_0  L^{\rm fast}  = \frac 12 (\dot A_j^a )^2 +  \frac m2 \epsilon _{jk} \dot A^a_j  A^a_k - 
\frac 12 (A^a_j A^a_k) (\vecb{C}^2 \delta_{jk} - C_jC_k) + {\rm other\ terms} \, ,
 \ee
 gives the QM propagator
\be
\label{GreenQM}
D_{jk}^{ab} = \frac {ig_0^2 \delta^{ab} } {\omega^2 - \vecb{C}^2 - m^2} \left[ \delta_{jk} - 
\frac {C_j C_k}{\omega^2} - \frac {im}\omega \epsilon_{jk} \right]
 \ee
(it is obtained from the field theory propagator (\ref{Green3d}) by replacing
$\vecb{p} \to \vecb{C}$, dividing by $L^2$ and assuming $a,b=2,3$).

We are hunting for the corrections $\sim \dot{C}_j {\cal A}_j(\vecb{C})$ in the effective Lagrangian.
\footnote{Do not confuse curly ${\cal A}_j$ with the physical gauge potentials $A_j$. 
The former are the functions of the latter !} 
To this end, we should pose
  \be
\label{Afon}
\vecb{C}(\tau) \ \to \ \vecb{C} + \vecb{E}\tau  \ , 
 \ee
( $\tau$ is Euclidean time; to evaluate the graphs, we are going to perform, as usual, the Wick rotation etc.)
and proceed in the same way as in the Appendix of the previous paper \cite{ja} where the calculation of the fermion
loop was described in details.

We just quote here the result of that calculation. The fermion loop contribution to the 
effective Lagrangian can be represented as
 \be
\Delta^F {L}_{\rm eff} \ =\ -E_j {\cal A}_j^F(\vecb{C}) 
 \ee 
with 
\be
\label{Aferm}
 {\cal A}_j^F (\vecb{C}) \ =\ 
 \frac {\epsilon_{jk} C_k}{2\vecb{C}^2} \left[ 1 -\frac {m} 
{\sqrt{m^2 + \vecb{C}^2}} \right] \ .
 \ee 
The corresponding magnetic field is 
\be
\label{Bferm}
{\cal B}^F = \epsilon_{jk} \partial_j {\cal A}_k^F  \ =\ -\frac {m}{2  (\vecb{C}^2 + m^2)^{3/2}}
 \ee
It has the flux
 \be
\label{fluxF}
q^F = \frac {\Phi^F}{2\pi} \ =\ \frac 1{2\pi} \int {\cal B}^F(\vecb{C}) \, d\vecb{C} = \ -\frac 12\ .
 \ee 

\vspace{1mm}

\begin{figure}[t]
\begin{center}
\includegraphics[width=3in]{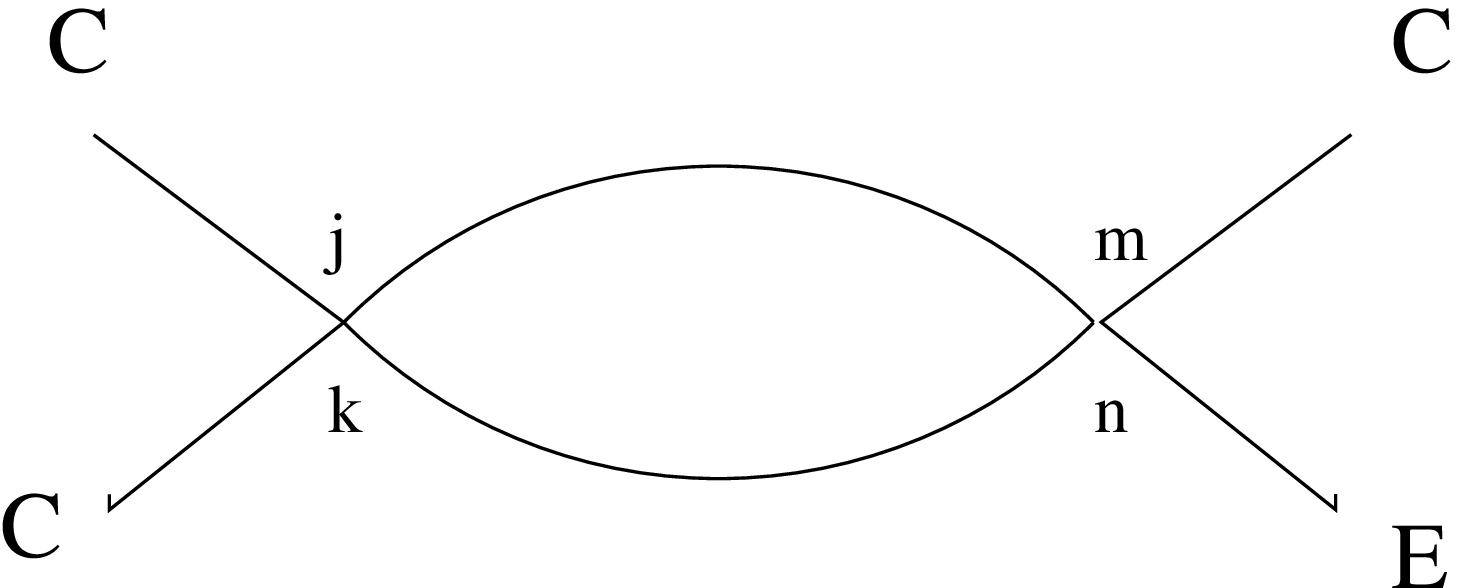}
\end{center}
\caption{}
\label{4tail}
\end{figure}

In the bosonic case, the lowest order (in the background) contribution is $\sim ECCC$ and is described 
by the graph in Fig. \ref{4tail} 

The calculation gives
\be
\label{Leff4tail}
{ L}_{\rm eff}^{\rm Fig.1} \ =\ \nonumber \\
\frac 1{4g_0^4} (\vecb{C}^2 \delta_{jk} - C_j C_k) 
[2(\vecb{C} \vecb{E}) \delta_{mn} - C_n E_m - C_m E_n]
\left. \int_{-\infty}^\infty \frac {d\omega}{2\pi} D_{jm}^{ab} (\omega)  \frac \partial {\partial \omega} D^{ab}_{nk} (\omega) \right |_{\vecb{C} = 0}
\nonumber \\ \  =
-2m  \vecb{C}^2 \epsilon_{jm} C_j E_m \int_{-\infty}^\infty \frac {d\Omega}{2\pi(\Omega^2 +  m^2)^3}  \ = \ - \frac 3{8 m^4} \vecb{C}^2
\epsilon_{jk} C_j E_k \ .
  \ee
The effective Lagrangian accepts also contribution $\sim ECCCCC$ from the six-leg graphs, etc. To sum them all up, one 
should (see Ref.\cite{ja} to understand why)
 \begin{itemize}
\item Write the expression (\ref{Leff4tail}) and restore the dependence on $\vecb{C}$ in Green's functions.
This gives a {\it phantasy} effective Lagrangian.
 \item  The {\it true} effective Lagrangian is obtained by multiplying the term 
$\propto E C^{2n+1}$ in the expansion of ${L}_{\rm eff}^{\rm phant}$
by $2/(n+1)$. 
\end{itemize}
 This finally gives
 \be
\label{Abos}
 {\cal A}_j^B(\vecb{C}) \ =\ -8m \, \epsilon_{jk} C_k \int_0^1 s^3 ds
\int \frac {d\omega}{2\pi(\omega^2 + s^2 \vecb{C}^2 + m^2)^3} = \nonumber \\
- \frac {\epsilon_{jk} C_k}{2\vecb{C}^2} \left[ 2 -\frac {3m} 
{\sqrt{m^2 + \vecb{C}^2}}  + \frac {m^3} {(m^2 + \vecb{C}^2)^{3/2}} \right] \ .
 \ee
The corresponding magnetic field is
 \be
\label{Bbos}
{\cal B}^B(\vecb{C}) \ =\ \frac {3m  \vecb{C}^2}{2  (\vecb{C}^2 + m^2)^{5/2}}
 \ .
 \ee
Its flux is
 \be
\label{fluxes}
q^B \ =\ 1 \   = \  -2 q^F \ .
 \ee

However, as was already emphasized in the Introduction, this effective action calculation cannot be trusted
because: 
\begin{itemize}

\item By its very meaning (relying on the smallness of fluctuations with respect to the background) , 
it makes sense only when $\vecb{C} \neq 0$.

 \item One can always add to the Lagrangian a total derivative. In our case, we can add a total derivative
{\it that is singular at $\vecb{C} = 0$}. In particular, one can add the derivative
$\frac {-i}2 \frac d{dt} \ln (C_+/C_-) $ which leads to the extra contribution 
 \be
\label{singA}
{\cal A}_j = \epsilon_{jk} \frac {C_k}{\vecb{C}^2}
 \ee
 in the effective vector potential. The calculation described above cannot ``detect'' this singular piece - 
by construction, the vector potential (\ref{Abos}) is analytic at $\vecb{C} = 0$.

The contribution (\ref{singA}) gives a delta-singular effective magnetic field with
the  flux $-1$, which would exactly cancel the gluon loop contribution.   

\end{itemize}

\section{Effective Hamiltonian.}
\setcounter{equation}0
 
The important message that we want to convey here  is that one {\it can} resolve this ambiguity, if using
the  Hamiltonian rather than Lagrangian language and matching the effective wave function on the Abelian moduli
space to the wave function in the vicinity of the origin. 
  
 We will proceed in the classical Born-Oppenheimer spirit,
subdivide all the variables into {\it slow} variables relevant to the low-energy dynamics and the {\it fast} ones
to be integrated over. Explicitly, we represent the full wave function as
 \be
\label{PsiBO}
\Psi_{\rm low\ energy}(x^{\rm fast}, x^{\rm slow}) \ \approx \ \chi^{\rm eff} (x^{\rm slow}) \Psi_0(x^{\rm fast})
\, ,
 \ee
with $  \Psi_0(x^{\rm fast})$ being the ground state of the fast Hamiltonian where the slow variables play the role of 
parameters. Then the effective Hamiltonian that acts on $\chi^{\rm eff}(x^{\rm slow})$  represents the average of the
full Hamiltonian over the fast vacuum state,
 \be
\label{Heffdef}
\hat H^{\rm eff} \ =\ 
\langle \hat H  \rangle_{\rm fast\ vacuum} \, .
 \ee

This method was used in \cite{Wit82} for non-chiral (3+1) supersymmetric gauge theories. The leading
order effective Hamiltonian describes in this case just the free motion over $T \times T \times T$ ($T$ being the maximal
torus of the group) with an additional Weyl invariance requirement imposed on the states. 
In \cite{chiralBO}, we applied this method for {\it chiral} (3+1) theories with left-right asymmetric
matter content. In this case, nontrivial PB phases appear. For example, in the chiral SQED with 8 left chiral
matter multiplets of charge 1 and a right chiral multiplet of charge 2, the motion runs over $T^3$ equipped
with a magnetic monopole of charge +7 and 7 monopoles of charge -1.   
 The method can be (and was \cite{popravki}) extended  such that loop corrections 
to any BO order in the effective Hamiltonian can be calculated,
but it suffices for us here to stay in the approximation (\ref{PsiBO}) and 
to evaluate (\ref{Heffdef}). 

For gauge theores, the Schr\"odinger equation should be supplemented by the Gauss law constraints. The latter
can be treated either as quantum constraints to be imposed on the states, or else one can resolve the constraints
at the classical level such that only gauge-invariant variables are left in the Hamiltonian. The former method is simpler
and we use it in the main text. For methodic purposes,
we repeated the analysis with the gauge constraints explicitly resolved, and this is the subject of Appendix D.
\footnote{For sure, such an analysis is not technicably possible in a gauge {\it field} theory. 
But in a gauge quantum 
{\it mechanics}, it is quite feasible.}

As was discussed above, we are basically interested only in the dynamics in one of the corners of the dual torus where
the extra contributions to the flux come from. In that case, higher Fourrier modes are irrelevant and we are in a position 
to study the dimensionally reduced SQM theory. The quantum supercharges in this reduced theory are 
\be
\label{QQbar}
g_0Q =  E^a_- \psi^a +  i  B^a \bar \psi^a \ , \ \ \ \ \ \ \ \ \   
g_0\bar Q =  \bar \psi^a E^a_+  - i  B^a  \psi^a\ ,
 \ee
where $g_0$ is the QM coupling constant (\ref{gQM}) of canonical dimension 
$m^{3/2}$  and 
  \be
\label{EPi}
E^a_j  = g_0^2\Pi^a_j - \frac m2 \epsilon_{jk}A^a_k
 \ee 
 with $\Pi^a_j = -i \partial/
\partial A^a_j$. The fermion variables are expressed
via the constant modes of the original
field theory variables in Eq.(\ref{LN1}) 
 as 
\be
\label{psilam}
 \psi^a = \frac {\lambda_1^{(\vecb{0})} - i \lambda_2^{(\vecb{0})} }{g_0\sqrt{2}}\, , \nonumber \\
 \bar \psi^a = \frac {\lambda_1^{(\vecb{0})} + i \lambda_2^{(\vecb{0})} }{g_0\sqrt{2}} \, .
 \ee
In holomorphic representation,  $\bar \psi^a \equiv \partial/\partial \psi^a$. Finally,
 \be
\label{B}
B^a = \frac 12 \epsilon^{abc} \epsilon_{jk} A^b_j A^c_k \ =  -\frac i2 \epsilon^{abc} A^b_-  A^c_+
 \ee
is the non-Abelian magnetic field strength. 
One can derive $Q^2 =  A^a_- G^a$, where
 \be
\label{Gauss}
G^a = \epsilon^{abc} (A^b_j \Pi^c_j - i \psi^b \bar \psi^c) 
 \ee
is the Gauss law. $Q$ and $\bar Q$ are nilpotent in the Hilbert space involving only  
gauge unvariant states. 
The anticommutator $\{Q, \bar Q \}/2$ gives the Hamiltonian,
 \be
\label{Ham}
 H = \frac {g_0^2} 2 \left(\Pi^a_j - \frac m{2g_0^2} \epsilon_{jk} A^a_k \right)^2 + \frac 1{4g_0^2} [(A^a_j A^a_j)^2 - A^a_j A^a_k A^b_j A^b_k]   \nonumber \\
+ i\frac {\epsilon^{abc} }2 \left[ \bar \psi^a \bar \psi^b A^c_+ +  \psi^a  \psi^b A^c_- \right]
 + \frac m2 \left( \psi^a \bar \psi^a - \bar\psi^a \psi^a \right) \ ,
 \ee
One can make three simple observations.
\begin{itemize}
\item The Hamiltonian (\ref{Ham}) involves two dimensionfull parameters, $g_0$ and $m$. 
They are ordered as
 \be
\label{ordergm}
 g_0^2 \ \gg m^3 \ .
 \ee
This is a corollary of the condition for the box to be small, $mL \ll 1$. As a result,
the mass terms in (\ref{Ham}) are smaller than the terms without mass. 
\item  The Hamiltonian  admits an integral of motion --- the angular momentum 
\be
\label{j}
j = \epsilon_{jk} A^a_j \Pi^a_k + \frac 12 \psi^a \bar \psi^a \, . 
 \ee
The eigenvalues of $j$ are integer for bosonic states and 
half-integer for fermionic states. Note that $j$ does not commute with the supercharges 
(\ref{QQbar}), such that a sector with definite $j$ is not supersymmetric. 
\item On the other hand, the Hamiltonian (\ref{Ham}) does not preserve the fermion charge. 
\footnote{It shares this feature with 
the Hamiltonian of  ${\cal N} = 4$  $4d$ SYM theory while, in  ${\cal N} = 1$ $4d$ theories,
the fermion charge is concerved \cite{Claud}.}
 That means that eigenfunctions of (\ref{Ham}) do not have a definite
fermion charge. The bosonic states represent a superposition of the terms of charge $F=0$ and $F=2$,
 \be
\label{Psibos}
\Psi \ =\ P +  \, \frac 12  \epsilon^{abc} (S A_-^a  + R A_+^a + T B^a  ) \psi^b \psi^c, 
 \ee 
with the scalar functions $P, S, R, T$  
depending only on three gauge-invariant variables ${\cal X} = A^a_+ A^a_-$, ${\cal Z} =   A^a_+ A^a_+$ and
$\bar {\cal Z} =   A^a_- A^a_-$. 
The wave functions (\ref{Psibos}), (\ref{Psiferm}) are the eigenstates of the operator (\ref{j}). This means that 
$P,S$ are transformed in the same way under rotations, while $T$ has the extra charge $-1$ and $R$ --- the extra charge
$-2$.

Likewise, the fermion states represent mixtures of the  $F=1$ and $F=3$ components,
\be
\label{Psiferm}
\Psi \ =\ \frac 16  P' \epsilon^{abc} \psi^a \psi^b \psi^c + 
( S' A_+^a  +  R' A_-^a +  T' B^a  )  \psi^a \ . 
 \ee
Again, $P'$ and $S'$ have the same charges, $T'$ has the extra charge $+1$ and $R'$ ---
the extra charge $+2$. 

\end{itemize}

To study the behavior of this system at the vicinity of the Abelian valley, 
it is convenient to subdivide  six bosonic variables $A^a_j$  into:
 \begin{itemize}
\item two Abelian slow variables $C_j \equiv A^1_j$,
\item   fast variables $A^{a=2,3}_j \equiv  b^a_j$ 
which in turn involve {\it i)} two projections
$b^a_j C_j$ --- the gauge degrees of freedom
 describing color rotations $A^1_j \to A^{2,3}_j$,
{\it ii)} After the partial gauge fixing
 $b^a_j C_j = 0$, we are left with two remaining  degrees of freedom, which
include    
 a physical gauge-invariant fast fluctuation variable $b^2 = (b^a_j)^2$ 
and an unfixed yet gauge angle describing the 
rotation around the first color axis 
(the chosen direction for the slow background).
\end{itemize}

The slow variables $C_j$ should lie in the range
\be
\label{rangeC}
g_0^{2/3} \ll   |\vecb{C}| \equiv a \ll \frac {g_0}{\sqrt{m}} \ .
 \ee
The lower bound here is the scale at which the characteristic values of $b$ becomes comparable to $a$
such that the BO approximation is no longer valid. The upper bound corresponds to $a \sim 1/L$.
 This corresponds to the interior of the dual torus where
 higher Fourrier harmonics (that we neglect) begin to play an important role.

In supersymmetric theory, the separation of fast and slow variables is more convenient 
to perform at the level of supercharges rather than for the Hamiltonian. The leading in the BO parameter
$\sim b/a$ part of the supercharges (\ref{QQbar}) can be represented as
  \be
\label{Qfast}
2 Q^{\rm fast} \  = \ \frac {g_0}{C_+} (C_+ \Pi^a_-  - C_- \Pi^a_+) \psi^a +  \frac  1{g_0} \epsilon^{ab} 
\bar \psi^a( C_+ b^b_-  -  C_- b^b_+) \ , \nonumber \\   
2\bar Q^{\rm fast} \  = \ \frac {g_0}{C_-} (C_- \Pi^a_+  - C_+ \Pi^a_-) \bar \psi^a +  
\frac  1{g_0} \epsilon^{ab}  \psi^a ( C_- b^b_+  -  C_+ b^b_-)  
 \ee
with $a,b = 2,3$. When deriving (\ref{Qfast}), we were allowed to replace
$$ \Pi^a_- \ \to \frac 12 \left( \Pi^a_-  - \frac {\Pi^a_+ C_-}{C_+} \right)\, , \ \ \ \ \ 
 \Pi^a_+ \ \to \frac 12 \left( \Pi^a_+  - \frac {\Pi^a_- C_+}{C_-} \right)\, , $$
bearing in mind that the Hilbert space where the fast 
supercharges act involves wave functions
 not depending on the projections $b^a_j C_j$ such that $C_j \Pi^a_j \Psi =  0$.
 If we also require the wave 
functions to be
annihilated by 
$\hat G^1 = \epsilon^{ab} ( b^a_j \Pi^b_j - i \psi^a \bar \psi^b )$, the 
 supercharges (\ref{Qfast}) become nilpotent. 

 The corresponding fast Hamiltonian is
\footnote{Cf. Refs. \cite{membKac} where a similar fast Hamiltonian for the 
quantum mechanics derived from N=4 $4d$ SYM theory 
was written and discussed.}  
 \be
\label{Hfast}
H^{\rm fast} \ =\ \frac {g_0^2}2 \Pi^a_+ \Pi^a_- - \frac 1{8g_0^2} \left( C_+ b^a_- - C_- b^a_+ \right)^2 
 + \frac i2 \epsilon^{ab} ( C_+ \bar \psi^a \bar \psi^b +
C_- \psi^a \psi^b ) \, .
 \ee

The Hamiltonian (\ref{Hfast}) represents a variety of supersymmetric oscillator.
It  has a single bosonic ground state. Up to a numerical factor, its wave function is
 \be
\label{Psifast}
\Psi^{\rm fast}_0 \ =\ (C_+ C_-)^{-1/4} \left[ 2i + \sqrt{\frac {C_-}{C_+}} \epsilon^{ab} \psi^a \psi^b \right] 
\exp \left\{  \frac 1 {8g_0^2 \sqrt{C_- C_+} } \left( C_+ b^a_- - C_- b^a_+ \right)^2  \right\}
 \ee

This is basically a product of Eq.(2.28) and Eq.(2.30) in Ref.\cite{ja}  with $m$ set to zero 
\footnote{It is more consistent {\it not}
to include the mass terms when writing the fast supercharges and the Hamiltonian, because  they are suppressed
compared to the others (see the comment after Eq.(\ref{Ham}) ).
 We emphasize that, while,
 in the Feynman graph method
addressed in the previous section, we were obliged to include the mass terms in the propagators 
to regularize  infrared singularities, we {\it do} not need to bother about mass in the Hamiltonian
approach.}
and generalized to an arbitrary   $C_j \neq C\delta_{j1}$. 
 We have included the factor $(C_- C_+)^{-1/4}$ in the definition of $\Psi^{\rm fast}_0$ for the normalization
integral 
 $$
\int dx^{\rm fast} \, \left| \Psi^{\rm fast}_0 \right|^2 
 $$
with 
\footnote{{\it This} particular form of $dx^{\rm fast}$, in particular the important 
factor $(C_+ C_-)^2$ there 
follows from the requirement that the original measure $\prod_{aj} dA^a_j$ goes over to 
$dC_+ dC_- dx^{\rm fast}$ on the valley. See also Eq.(\ref{measureabalph}).}
\be
\label{merafast}
dx^{\rm fast} \sim \ (C_+ C_-)^2  d_{\rm fermions} \,d^2 b_+ d^2 b_- \, \prod_{a=2,3} \delta(C_+ b^a_- + C_+ b^a_+)
 \ee
 not to depend on $C_j$. 

Note that the fermion factor in (\ref{Psifast}) involves only two terms, not four terms as in  
 a generic decomposition (\ref{Psibos}). That is  because, {\it at the valley}, there is no difference between
3 bifermion structures in (\ref{Psibos}). They are expressed into one another by multiplying over a proper function
of slow variables $C_\pm$.

With the fast ground state wave function in hand, we can determine the effective Hamiltonian. 
Assume first $k > 0$. By analyzing the effective Hamiltonian in the interior of the dual
torus, we have seen that the vacuum states are in this case bosonic (see Eq.(\ref{Psim})). 
This should concern also the
Hamiltonian (\ref{Ham}) describing the ``corner dynamics''.  Thus, the vacuum wave function annihilated
by the full supercharges (\ref{QQbar}) has the form (\ref{Psibos}). Consider this function and the
equations $\hat Q \Psi = \hat {\bar Q} \Psi = 0$ near the Abelian valley (\ref{rangeC}). The wave 
function there is approximately
given by the product [cf. Eq.(\ref{PsiBO})]
 \be
\label{PsiBvalley}
 \Psi_0 \ =\ \chi^{\rm eff} (C_j) \Psi_0^{\rm fast} 
 \ee
with $ \Psi_0^{\rm fast} $ written in (\ref{Psifast}).  
The effective supercharges acting on $\chi^{\rm eff} (C_j)$
are 
 \be
\label{Qeffdef}
Q^{\rm eff} = \langle \Delta Q \rangle_0 \, , \ \ \ \ \ \ \ \ 
 \bar Q^{\rm eff} = \langle \Delta \bar Q \rangle_0
  \ee
with 
 \be
\Delta Q = Q - Q^{\rm fast} = 
-i\psi^1 \left( 2g_0 \frac \partial {\partial C_+} + \frac m{2g_0} C_- \right) + \frac {\bar \psi^1}{2g_0}
\epsilon^{ab} b^a_- b^b_+ - \frac {im}{2g_0} \psi^a b^a_- \, , \nonumber \\
\Delta \bar Q = \bar Q - \bar Q^{\rm fast} = 
- i\bar\psi^1 \left( 2g_0 \frac \partial {\partial C_-} - \frac m{2g_0} C_+ \right) - \frac {\psi^1}{2g_0}
\epsilon^{ab} b^a_- b^b_+ + \frac {im}{2g_0} \bar\psi^a b^a_+ \,   .
 \ee
Two last terms give zero after averaging. We obtain
   \be
\label{Qeff}
Q^{\rm eff} =  
-i\psi^1 \left( 2g_0 \frac \partial {\partial C_+} + \frac m{2g_0} C_- \right) - 2ig_0 \psi^1 \left\langle \frac
\partial {\partial C_+} \right \rangle_0 \, , \nonumber \\
 \bar Q^{\rm eff} =  
-i\bar\psi^1 \left( 2g_0 \frac \partial {\partial C_-} - \frac m{2g_0} C_+ \right) - 2ig_0 \bar \psi^1 \left\langle \frac
\partial {\partial C_-} \right \rangle_0 \, .
 \ee
The last terms above involve PB phases,
 \be
\label{Berry+-}
{\cal A}_-^{\rm PB} \propto \label{calA+-}
\left\langle \frac
\partial {\partial C_+} \right \rangle_0  \ =\ - \frac 1 {4C_+}, \ \ \ \ \ \ \ \ \ 
 {\cal A}_+^{\rm PB} \propto    \left\langle \frac
\partial {\partial C_-} \right \rangle_0  \ =\  \frac 1 {4C_-}
 \ee
and hence
 \be
\label{Berryj}
 {\cal A}_j^{\rm PB} \propto \frac {\epsilon_{jk} C_k}{\vecb{C}^2} \ .
 \ee
Note that the averages (\ref{calA+-}) depend only on the fermion factor in the fast wave function 
(\ref{Psifast}), the bosonic factor does not produce any phases. Indeed, nontrivial PB phases
 \be
\label{Berry}
{\cal A}_j^{\rm PB} \propto \frac {\int (\Psi_0^{\rm fast})^* \frac {\partial }{\partial C_j} \Psi_0^{\rm fast} \, dx^{\rm fast} }
{\int (\Psi_0^{\rm fast})^* \Psi_0^{\rm fast} \, dx^{\rm fast} }
   \ee
can arise only due to complexities in the wave function. The {\it real} bosonic exponential factor 
{\it could} contribute a total gradient in ${\cal A}_j^{\rm PB}$ ( which could then be eliminated by  a gauge transformation), but
this gradient is just absent, if choosing the normalization factor as in (\ref{Psifast}). 

  Thus, the effective supercharges involve tree level vector potentials ${\cal A}^{\rm tree}_\pm \propto \pm kC_\pm$ (these
are vector potentials for the constant magnetic field on the dual torus of total flux $2k$) and the induced potentials
(\ref{Berry+-}). The latter have the form (\ref{singA}) with the factor $1/2$. 
These potentials have a delta-functional magnetic field.
\footnote{This is so for zero mass. When $m \neq 0$, the flux is concentrated
in the region $a \sim m$, which is much smaller than the lower bound in
(\ref{rangeC}).}
One can be convinced that the corresponding flux is equal to $-1/2$, which coincides with the flux brought about by
the {\it fermion} loops in the effective action method. On the other hand, there is no trace of the gluon loop contribution 
in the Hamiltonian approach !

Consider now the case $k < 0$. The ground states are now fermionic having the form (\ref{Psiferm}). 
In the vicinity of the valley, the wave function is natural to represent as
 \be
\label{PsiFvalley}
 \Psi_0 \ =\ \psi^1\chi^{\rm eff} (C_j) \tilde \Psi_0^{\rm fast}
 \ee
with
  \be 
\label{dvefast} 
 \tilde \Psi_0^{\rm fast} \ =\ \sqrt{ \frac {C_+}{C_-}} \Psi_0^{\rm fast} \ .
 \ee

The appearance of the extra factor $ \sqrt{ {C_+}/{C_-}}$ in (\ref{dvefast}) reflects the presence of the factor
$A^a_+$ in the second term in (\ref{Psiferm}) rather than the factor $A^a_-$ in the second term in (\ref{Psibos}).
(As was discussed above, in the vicinity of Abelian valley, two other fermion bilinear terms are reduced to the term 
$\propto S A^a_-$ in the bosonic case and to the term $\propto S' A^a_+$ in the fermionic case. For less heuristic justification of the choice
(\ref{dvefast}), see the footnote after (\ref{obreshchi}) below.)

When transferring the analysis above to the case $k < 0$ with the modified fast vacuum function (\ref{dvefast}), we obtain
the induced singular vector potentials like in (\ref{singA}) with the {\it positive} $\delta$-functional flux 
$\Delta \Phi_{k < 0}/(2\pi)
 = 1/2$ ( the contribution $-1/2$ coming from $\Psi_0^{\rm fast}$, as above and the contribution $+1$ from the
factor $\sqrt{ {C_+}/{C_-}}$. 

The sign of the induced flux is thus always opposite to the tree-level flux. Recalling that the full dual torus includes 
four  singular points, one obtains the flux renormalization $2k \to 2(k-1)$ in the case $k > 0$ and 
$2k \to 2(k+1)$ in the case $k < 0$. This finally gives the answer (\ref{Ik2}) for the
index.     

At this stage, our findings have the flavour of a paradox. Indeed, we discovered in
the previous section that the effective Lagrangian calculations involve an intrinsic ambiguity
associated with adding a singular total derivative. Now we are claiming 
to resolve this ambiguity in the Hamiltonian approach. 
But the Lagrangian and Hamiltonian descriptions must be completely equivalent. How come ?

The resolution of this paradox is the subject of the next section.

\section{Torus with the corners.}
\setcounter{equation}0

Note first of all that there {\it is} an ambiguity also in the Hamiltonian method that exactly
corresponds to the Lagrangian ambiguity mentioned above.
 One is always allowed to introduce a factor
$( {C_+}/{C_-})^\alpha $ in the fast vacuum wave function --- 
the slow variables $C_j$ enter as parameters 
in the fast Hamiltonian and we cannot decide whether to include 
this factor in the definition of  $\Psi_0^{\rm fast}$
or not. This uncertainty translates into the uncertainty of the coefficient of the 
$\delta$-functional flux
 located at the points where the BO approximation breaks down. 
\footnote{The ambiguity of this kind can appear only in a (2+1)-dimensional problem. In chiral
(3+1)-dimensional theories, PB phases are unambigously fixed 
(up to a gauge
transformation) \cite{chiralBO}. Indeed, the induced field
there has not the form of flux lines, but rather of a 
magnetic monopoles with a nonzero magnetic 
field strength not only at the origin, but also in its vicinity. It cannot be mimicked neither by a total derivative
in the effective Lagrangian, nor by a factor entering the definition of the fast ground state wave function.}

In particular, choosing
$\alpha = 1/2$ effectively brings about the additional unit flux in the origin. In the full
problem, this amounts to adding four units of flux (one in each corner), which exactly 
imitates the gluon loop contribution. 

This ambiguity cannot be resolved while staying on the Abelian valley. Our main point is,
however, that it {\it can} be fixed if imposing the {\it additional} requirement for the wave
function of the full QM Hamiltonian (\ref{Ham}) to be regular at the origin $A_j^a = 0$.

Unfortunately, right near the corner where the Abelian BO approximation breaks down,
the equation $Q\Psi = 0$ cannot be solved analytically. Still, 
 we can  
approach the corner from  the Abelian valley side. Consider the effective supercharge
(\ref{Qeff}). Bearing in mind (\ref{Berry+-}), it is proportional to 
 \be
\label{Qeffcanon}
Q^{\rm eff} \ \propto \ \frac \partial {\partial C_+} - \frac 1{4C_+} + \frac m{4g_0^2} C_- \ .
 \ee 
 A generic solution to the equation $Q^{\rm eff} \chi(C_k) = 0$ is
\be
\label{obreshchi}
 \chi(C_k) \sim (C_- C_+)^{1/4} P(C_-) \exp \left\{    - \frac {m C_- C_+}{4g_0^2}\right\}
 \ee
with an arbitrary entire function $P(C_-)$. 
In the range (\ref{rangeC}), the exponential factor in Eq.(\ref{obreshchi}) is close to 1 and irrelevant.
\footnote{
When $k <0$, the effective wave function involves the factor $\psi^1$. It is annihilated automatically by $Q$, while
\be
\label{Qbareffcanon}
\bar Q^{\rm eff} \ \propto \ \frac \partial {\partial C_-} - \frac 1{4C_-}  \ 
 \ee 
and hence 
\be
\label{obreshchiF}
 \chi(C_k) \sim \psi^1 (C_- C_+)^{1/4} P(C_+)
 \ee
{\it if the fast wave function is chosen as in (\ref{dvefast}) }. An eigenstate with a definite (half-integer in this case) 
$j$ behaves at the origin 
as 
\be
\label{chieffF}
\chi_j(C_k) \ \sim \ \psi^1 (C_-C_+)^{1/4} C_+^{j-1/2}  \ ,
 \ee
which is regular when $j \geq 1/2$.}
An eigenstate with a definite
angular momentum (\ref{j}), or rather its effective counterpart in the sector $F=0$,
 \be
\label{jeff}
\hat j ^{\rm eff} \ =\ -i\epsilon_{jk} C_j \frac \partial {\partial C_k} \, ,
 \ee
behaves at the origin as 
 \be
\label{chieff}
\chi_j(C_k) \ \sim \ (C_-C_+)^{1/4} C_-^{-j}  \ .
 \ee
If we require for  $\chi_j(C_k)$  to be nonsingular at the origin, $j$ must be negative integer or zero. A glance
at (\ref{Psifast}) tells us that, for $j \leq 0$, also the 
 full wave function 
(\ref{PsiBvalley}) behaves as 
 \be
\label{Psi0a-j}
\Psi_0 \sim a^{-j}
 \ee
 and is nonsingular. And, for positive $j$, it is singular.
Such solutions should be excluded. 
 \footnote{
 The representation (\ref{PsiBvalley}) holds only  in the valley approximation and, strictly speaking,
 we are  not allowed
to go with it right into the origin. 
It happens, however, that, irrespectively of whether the Abelian BO approximation is valid or not, 
one {\it can} follow
the Abelian valley up to the very origin and rigourously {\it prove} that nonsingular
at the origin wave function {\it excludes} positive $j$. This proof is the subject of Appendix C.} 

Suppose now that we redefined the fast wave function according to
 \be
\label{fastwithfactor}
\tilde \Psi_0^{\rm fast}  \ =\ \sqrt{\frac {C_+} { C_-} }  \Psi_0^{\rm fast}\ .
 \ee
The effective supercharge behaves now at the origin as
  \be
\label{Qefftilde}
\tilde Q^{\rm eff} \ \propto \ \frac \partial {\partial C_+} + \frac 1{4C_+} \ .
 \ee   
The effective angular momentum operator is also modified,
  \be
\label{jefftilde}
\hat j ^{\rm eff} \ =\ -i\epsilon_{jk} C_j \frac \partial {\partial C_k} + 1\, .
 \ee 
An eigenfunction of (\ref{Qefftilde}) with a definite value of $j$ is 
 \be
\label{chiefftilde}
\tilde \chi_j(C_k) \ \sim \ (C_-C_+)^{-1/4} C_-^{1-j}  \ .
 \ee
Again, for $\tilde \chi(C_k)$ to be nonsingular, the condition $j \leq 0$ should be satisfied.
When multiplied by $\tilde \Psi^{\rm fast}_0$ with the factor $\sqrt{C_+/C_-}$ it now includes, we obtain
{\it the same} full wave function (\ref{PsiBvalley}) as before.

In other words, it does not matter at all whether the contribution of gluon loops is
taken into account or not. What {\it is} important is to pose proper boundary conditions 
in the corners of the torus. Having done that, we are able to count the states and evaluate the index.

Let us first remind how it is done at the tree level without yet taking any loop (fermion or gluon) into account.

The equation $Q^{\rm eff} \chi = 0$  boils down in this case to
 \be
\label{QeffPsi} 
 \left( \frac \partial {\partial z} + \pi k \bar z \right) \chi \ =\ 0
 \ee
($z = x+ iy$ with $x,y$ being defined in (\ref{xy}) ). A generic solution to Eq.(\ref{QeffPsi}) is
 \be
\label{solFzbar}
 \chi \ =\ e^{-\pi k \bar z z} F(\bar z) \, ,
 \ee 
where $F(\bar z)$ is any antiholomorphic function. The particular solutions (\ref{Psim}) are obtained if imposing 
proper boundary conditions \cite{Deser,ja} 
\be
\label{bc2}
\chi(x+1, y) &=& e^{-2\pi i ky} \chi(x,y) \ , \nonumber \\
\chi(x, y+1) &=& e^{2\pi i kx} \chi(x,y) \ .
 \ee
They can  be represented as
 \be
\label{chim}
 \chi_m = e^{-\pi k \bar z z} e^{\pi k \bar z^2} Q^{2k}_m(\bar z)
 \ee
 in the notations of Appendix A. A kinship of the wave functions (\ref{chim}) to the wave
functions (\ref{psiQ}) of the pure CS states is clearly seen. It is the same kinship as between the wave functions
of the lowest Landau levels and the wave functions of the states in the topological theory with the Lagrangian
$\sim B \epsilon_{jk} \dot x_j x_k $.  

Now, take loops into account. 
Call for definiteness $\chi(C_k)$ the coefficient of the fast wave function
(\ref{Psifast}). Then, as we have seen, singular fluxes $\Phi/(2\pi) = -1/2$ are added in each corner. 
The gauge field due to each such flux line is a  singular  
pure gauge, like in Eq.(\ref{singA}). 
This brings about a factor $\sim (C_+/C_-)^{1/4} \ \to \ (z/\bar z)^{1/4}$ in the corner $z=0$ and, 
similarly, in the other corners. Thus, the effective
nonsingular wave functions satisfying the boundary conditions (\ref{bc2}) have the form
 \be
\label{Psiwithfluxes}
 \chi_m  =  e^{-\pi k \bar z z + \pi k \bar z^2} \prod_{np} 
\left( \frac {z+n/2 + ip/2}{\bar z+n/2 -ip/2 } \right)^{1/4}
Q^{2k-2}_m (\bar z) \sqrt {Q^4_3(\bar z) -  Q^4_1(\bar z) } \, ,
 \ee
where the product runs over all integer $n,p$. The argument of the square root is the function
(\ref{Pi}) with four zeros in the four corners.
The square root  has branching points at the corners, but the 
full functions (\ref{Psiwithfluxes}) are regular there. 
Note that $m$ runs now from 0 to $2k-2$, 
which gives finally $k$ (rather than $k+1$) solutions in accordance with (\ref{Ik2}).    

\vspace{2mm}

\centerline{\large \sl $\sigma$-model on the quotient.}

\vspace{2mm}

When counting the states, we {\it first} have found  all the regular
solutions of (\ref{bc2}) for the functions having the form 
(\ref{solFzbar})  (when staying at the tree level) or involving extra $z$-dependent factors as in 
(\ref{Psiwithfluxes}) (when extra fluxes at the corners are taken into account).
{\it Then} we imposed the Weyl invariance requirement.

Another way to handle this problem  is to factorize our torus over the Weyl group
and study the effective theory on the quotient \cite{Wit99}. For $SU(2)$, the Weyl group is just $Z_2$. As is well known,
$T^2/Z_2 = S^2$
\footnote{One of the many ways to see it is meditating over Fig.5 of Ref.\cite{ja}. Note also 
that, for $SU(N)$, the corresponding quotient is $[T^{\rm max} \times T^{\rm max}]/S_N = \mathbb{CP}^{N-1}$ 
\cite{FMW}.}
Thus, the effective theory with all gauge constraints resolved represents a certain 
$\sigma$-model on $S^2$. What particular model is it ?

At the level of $T^2$, the effective supercharges were evaluated to have the form (\ref{Qeffcanon}).
A mathematician would call this differential operator a twisted antiholomorphic derivative ({\it twisting}
means adding an Abelian gauge field). The presence of extra Grassmann factor in $Q^{\rm eff}$ promotes
it to the twisted antiholomorphic {\it exterior} derivative. When going down onto the quotient, 
the supercharges should keep this form. 

We are thus arriving at the twisted Dolbeault complex. The  twist (e.g. the magnetic flux or the second
Chern class of the gauge field) is a half of the twist on $T^2$. When extra fluxes due to fermion loops 
are taken into account, we obtain the twist $(2k-2)/2 = k-1$. 
It is rather remarkable that this twisted Dolbeault
complex is equivalent to the {\it Dirac} complex for the field of flux $k$. 
\footnote{A mathematician can consult e.g. the Propositions 1.4.23 and 1.4.25 in the book \cite{Nicola} and a
 physicist may look into \cite{IvSm} for pedagogical explanations.} The Dirac index on $S^2$ is equal to $k$.

Incorrect results could be obtained if
 \begin{enumerate}
\item Not taking into account extra fluxes. This would give  twist $k$ for the Dolbeault complex
and  twist $k+1$ for the Dirac complex. This is the number of states in pure CS theory.
 \item Taking into account {\it both} fermion-induced and gluon-induced fluxes as in \cite{ja}. This would give
$k+1$ for the Dolbeault twist and $k+2$ for the Dirac twist.
\end{enumerate} 

Also for other unitary groups, the index (\ref{IkN}) coincides with the Dirac index on  $\mathbb{CP}^{N-1}$ with a {\it properly} chosen
gauge field. Adding gluon-induced fluxes would amount to the shift $k \to k+N$. 
If {\it no} extra fluxes were added, we would
obtain the tree level result  
\be
\label{IkNtree}
I(k,N) \ =\  
[{\rm sgn}(k)]^{N-1} \left( \begin{array}{c} |k|+N -1 \\ N-1 \end{array} \right)\ .
 \ee
which would make sense, for odd $N$, not for half-integer values of $k$, as it should \cite{Kao}, but
for integer ones.

Our final remark is that, though the reduction of a complicated field theory problem to a much simpler problem 
of calculating the Dirac index on  $\mathbb{CP}^{N-1}$ looks as a nice Christmas gift,  we do not see 
any other way to {\it prove} that the corners contribute the flux that exactly compensates
the flux associated with the square root of the canonical bundle (the difference between the 
Dirac twist and the Dolbeault twist)  than to perform an accurate effective Hamiltonian analysis as
we did in Sect. 3 and Sect. 4 above and in Appendix C below.

\section*{Acknowledgements}
I am indebted to E. Witten for many illuminating discussions and correspondence and to M. Konyushikhin
for reading the manuscript and useful comments.

\section*{Appendix A. Theta functions.}
\setcounter{equation}0
\def\theequation{A.\arabic{equation}} 
\vspace{1mm}

We remind here certain mathematical facts 
concerning the properties of analytical functions 
on the torus. They are mostly taken from the textbook \cite{Mumford}, but we are using different notations
which we find more clear and more appropriate for our purposes.

Theta functions play the same role for the torus as ordinary polynomials for the Riemann sphere. 
They are analytic, but satisfy certain nontrivial  quasiperiodic boundary conditions with respect to shifts
 along the cycles of the torus. 
A generic torus is characterized by a complex modular parameter $\tau$, but we will stick to the 
simplest choice $\tau = i$ so that the torus represents a square $x,y \in [0,1]$ ( $ z = x+iy$) glued around.

The simplest $\theta$-function satisfies the boundary conditions
 \be
\label{thet1bc}
 \theta(z+1) &=& \theta(z) \, ,\nonumber \\
 \theta(z+i) &=& e^{ \pi (1 - 2iz)} \theta(z) \, .
 \ee
This defines a {\it unique} (up to a constant complex factor) analytic function. Its explicit
form is
 \be
\label{thet1}
\theta(z) \ =\ \sum_{n = -\infty}^\infty \, \exp\{- \pi n^2 + 2\pi i n z \} \, .
 \ee
This function (call it theta function of level 1 and introduce an alternative
notation $\theta(z) \equiv Q^1(z)$) has only one zero in the square $x,y \in [0,1]$ --- 
right in its middle, 
$\theta(\frac {1+i}2 ) = 0$. For any integer $q > 0$, one can define theta functions
of level $q$ satisfying
 \be
\label{thetqbc}
 Q^q(z+1) &=& Q^q(z) \, ,\nonumber \\
 Q^q(z+i) &=& e^{q \pi  (1 - 2iz)} Q^q(z) \, .
 \ee
When $q >1$, the functions satisfying (\ref{thetqbc}) lie in  vector space of dimension $q$. The basis
in this vector space can be chosen as 
 \be
\label{Qqm}
Q^q_m(z) \ =\ \sum_{n = -\infty}^\infty \, \exp\left\{- \pi q \left(n + \frac mq \right)^2 + 2\pi i q z 
\left( n + \frac mq \right)\right\} \, , \ \  \ \ m = 0, \ldots, q-1 \, .
 \ee
$Q^q_m(z)$ can be expressed in the notation of \cite{Mumford} as
  \be 
\label{QviaMum}
 Q^q_m (z) \ =\ \theta_{m/q,0} (qz, iq) \ ,
 \ee
where $\theta_{a,b}(z, \tau)$ are theta functions of rational characteristics. 

$Q^q_m(z)$ can be called  ``elliptic polynomials'' of order $q$. Indeed, each $Q^q_m(z)$ 
has $q$ simple zeros at
 \be
\label{zs}
z_s^{(m)} \ =\ \frac {2s+1}{2q} + i \left(  \frac 12   - \frac mq \right)\, , \ \ \ \ \ \ \ 
s= 0,\ldots,q-1 
 \ee
(add $i$ to bring it onto fundamental domain $x,y \in [0,1]$ when necessary). 
A product $Q^q(z) Q^{q'}(z)$ of two such ``polynomials'' of orders $q, q'$ gives a polynomial of order
 $q + q'$. 

For example,
$\theta^2(z)$, having the zero of order 2, can be represented as a superposition
 \be
\label{thetakvadrat}
 \theta^2(z) \ =\ \alpha Q^2_0(z) + \beta Q^2_1(z) \, .
 \ee
The coefficients $\alpha, \beta$ can be determined. A great number of similar relations between
theta functions of different levels can be written. We can amuse the reader with a relation
 \be
\label{642}
\frac {Q^6_5 - Q^6_1}{(Q^4_3- Q^4_1) Q^2_0}  \ =\ \frac 1{\eta(i)} = 
\frac {2\pi^{3/4}}{\Gamma(1/4)}
 \ee
with some physical implications to be discussed soon.

A ratio of different theta functions of the same order
 \be
\label{R}
R(z) \ =\ \frac {Q^q(z) }{\tilde Q^q(z)}
 \ee
represents a periodic meromorphic function. A properly defined
number of zeros of this function (such that a zero of the second order
is counted twice, etc) coincides with the properly defined number of its poles
(the {\it Riemann-Roch} theorem).

\section*{Appendix B. Counting of states in pure CS theory.}
\setcounter{equation}0
\def\theequation{B.\arabic{equation}}

 We just outline here the main steps of the analysis of Refs.\cite{Eli}. 
A reader is invited to look into the original papers and into the 
review \cite{Dunne} for more details.

The first remark is that the pure Chern-Simons is a topological theory 
involving zero Hamiltonian and a finite number
of states. Their wave functions depend not on both $A_1$ and $A_2$ as is
the case in the dynamical YM-CS theory, but rather on the
antiholomorphic combination 
$\bar A = A_- = A_1 - i A_2$, with  $A_+ = A_1 + i A_2$
playing the role of canonical momenta.
\footnote{The reason by which the wave functions should be chosen to be 
antiholomorphic 
is explained in the paragraph after
Eq.(\ref{Pi}). Note that, in Refs.\cite{Eli,Dunne}
 the wave functions  depended on $A_+$ rather than $A_-$ 
due to a different sign convention for $k$.}

We put the theory on the spatial torus of size $L=1$ (as this theory
does not involve dimensional constants, we cannot say whether the volume 
is large or small and will measure everything in the units of $L$).

A generic couple of matrix-valued Hermitian fields 
$A_j$  on the torus can be parametrized as \cite{Yang}
 \be
\label{AviaU}
 A_1 - iA_2 \ =\ 2\pi U(\vecg{\xi})  \bar z \sigma^3 U^{-1}(\vecg{\xi}) - 
i \partial_- U(\vecg{\xi}) \cdot U^{-1}(\vecg{\xi}) \, ,
 \ee
where $\xi_{1,2}$ are physical spatial coordinates, 
$\bar z=x-iy$ is a constant complex number.
$U = \exp\{(i\alpha^a - \beta^a) \sigma^a\}$ is a $SL(2,C\!\!\!\!C)$ matrix.  
When $\beta^a = 0$, $U \in SU(2)$ and (\ref{AviaU}) is reduced to a gauge
transformed constant field. In this case (but not in a generic case), the conjugated field
$A_+$ can also be represented as in (\ref{AviaU}) with $\partial_+$ being replaced for  
$\partial_-$ --- see Eq.(\ref{Ba+}) below.  

The wave functions must satisfy the Gauss law constraints. 
In the pure CS case, they boil down to  
 \be
\label{GaussCS} 
F_{12} \Psi[\bar A] = 0.
 \ee
 The solution
to these constraints is 
 \be
\label{PsiA}
\Psi[\bar A] \ =\ \psi(\bar z) \, \exp \left\{-k S_+[U] - \frac 
{ik \bar z}2 \int \langle 
\sigma^3 U^{-1} \partial_+ U \rangle d^2\xi \right\}\, ,
 \ee
 with $S_+[U]$ being  the Polyakov-Wiegmann functional \cite{PW},
 \be
\label{S+U}
 S_+[U] \ =\ \frac 1{8\pi} \int_{T^2} \langle U^{-1} \partial_- U U^{-1}  \partial_+ U \rangle 
+ \frac i{12 \pi} \int_{(3)} \epsilon^{\mu\nu\rho} \langle U^{-1}  \partial_\mu U  
U^{-1}  \partial_\nu U  U^{-1}  \partial_\rho U \rangle \, .
 \ee
The integral in the second term runs over a 3-manifold with the border $T^2$ and $\langle 
\cdots \rangle$ stands for the trace. 

To check the validity of (\ref{GaussCS}), one should act on the wave function (\ref{PsiA}) by the operator
 \be 
 A^a_+ = \ \frac 2\kappa \left( \frac \delta {\delta A^a_1} + i  \frac \delta {\delta A^a_2} \right) 
 = \ \frac 4\kappa \frac \delta {\delta A^a_-} 
 \ee
and be convinced (see \cite{Eli,Dunne} for details) that one thus obtains a factor
 \be
\label{Ba+} 
  B^a_+ \ =\ 2\pi U  \bar z \sigma^3 U^{-1} - 
i \partial_+ U \cdot U^{-1} 
 \ee
in front of $\Psi[\bar A]$, as if it were a pure gauge transformation. The differential
operator $F_{12} = (i/2)F_{+-}$ gives then $\partial_+ A_- - \partial_- B_+ - i[B_+, A_-]$, which is zero. 

On top of (\ref{GaussCS}), one should require the wave functional (\ref{PsiA}) to be invariant with respect
to two {\it large} (uncontractable) gauge transformations with the matrices
\be
\label{largegauge}
U_1(\vecb{\xi})  = e^{2\pi i \xi_1 \sigma_3}, \ \ \ \ \ \ \ \ \ 
U_2(\vecb{\xi}) = e^{-2\pi i \xi_2 \sigma_3}
 \ee
They correspond to the shifts $\bar z \to \bar z+1$ and $\bar z \to \bar z+i$. This brings $z$ onto the dual
torus, $x,y \in [0,1]$. The invariance under (\ref{largegauge}) implies the boundary 
conditions
 \be
\label{bcpsi}
\psi(\bar z+1) = e^{\pi k(1+ 2\bar z) } \psi(\bar z) \, , \nn 
\psi(\bar z+i) = e^{\pi k(1- 2i\bar z) } \psi(\bar z) \, .
 \ee
And that means that 
 \be
\label{psiQ}
 \psi(\bar z) \ =\ e^{\pi k \bar z^2} Q^{2k} (\bar z) \, ,
 \ee
where $ Q^{2k} (z)$ is a theta function of level $2k$. 

Finally, we impose the requirement of Weyl invariance, $\psi(-\bar z) = \psi(\bar z)$. This reduces
the number of states from $2k$ (the dimension of the vector space of $ Q^{2k} (\bar z)$)
down to $k+1$. This gives
 \be
\label{chislo}
 \#_{\rm states} ({{\rm pure} \ CS, \ SU(2)}) \ =\ k+1 \, .
 \ee
Wave functions (\ref{PsiA}) can in principle
 be used to calculate certain averages, e.g. the Wilson loop averages related
to knot invariants \cite{WitCMP}.
\footnote{We are not aware of such a direct calculation, however.}
To this end, one should know the functional integral measure ${\cal D}A$.
This measure was calculated in \cite{Gawed} with the result
 \be
\label{measGawed}
\langle \psi_1 | \psi_2 \rangle \ =\ \int dz d\bar z \, \psi_1^*(z)\, \psi_2(\bar z) 
e^{-2\pi (k+2) z \bar z} |\Pi(z)|^2 \, ,
 \ee
where $\Pi(z)$ is  a certain theta function
of level 4 having zeros at the ``corners'' of the dual torus, $\bar z = 0, 1/2, i/2, (1+i)/2$. In our notations,
\be
\label{Pi}
\Pi(z) = 
Q^4_3(z) - Q^4_1(z) \, .
 \ee    
It is {\it antisymmetric} in $z$. 

Note that the measure involves the exponential factor 
$\exp\{-2\pi (k+2) \bar z z \}$ which makes the integral convergent at large
$|z|$. The positivity of the exponent there is due to the fact that our 
wave functions were chosen to be antiholomorphic. Holomorphic functions would
lead to an inadmissible measure $\sim \exp\{2\pi (k+2) \bar z z \}$.

For some purposes, it might be convenient to represent the Weyl invariant
wave functions $Q^{2k}(\bar z)$ as a {\it ratio}
 of Weyl-antiinvariant functions 
of level $2(k+2)$ and the Weyl-antiinvariant function $\Pi(\bar z)$, like in
 (\ref{642}). Obviously, there are
$(k+2)-1$ Weyl-antiinvariant functions $ Q^{2(k+2)}(\bar z)$, the number coinciding with (\ref{chislo}).
This works also for all other groups. The number of states can be counted as the number of 
generalized Weyl-invariant functions characterized  by the integer $k$ or else as the number of 
 Weyl-antiinvariant functions characterized by the integer $k+h$, where $h$ is the dual Coxeter
number. However, if we are interested only in the state counting (as we are in this paper),
 and not in 
calculating averages, etc, the existence of the map $Q^{2k}_{\rm Weyl\ inv.} \to
Q^{2k+4}_{\rm Weyl\ antiinv.}$ is irrelevant.

This was all done for positive $k$. For negative $k$, wave functions depend on $A_+$ rather than on 
$A_-$ (such that the exponential factor in the measure provides, again, a suppression at large $|z|$), 
but this is the only change. The whole analysis can be 
repeated with the result $|k| + 1$ for the number of states.

\section*{Appendix C. Wave function at the origin.}
\setcounter{equation}0
\def\theequation{C.\arabic{equation}}

We will analyse here  the ground states of the SQM Hamiltonian (\ref{Ham}) and prove that, when $k > 0$,
  the states
with positive eigenvalues of the momentum (\ref{j}) are necessarily singular at the origin and should be
excluded from the spectrum. The case $k <0$ can be treated similarly, then negative $j$ are excluded.

A generic gauge-invariant bosonic wave function is written in Eq.(\ref{Psibos}). The functions $P,S,R,T$ depend
on three gauge-invariant variables ${\cal X} = A^a_+ A^a_-$, ${\cal Z} =   A^a_+ A^a_+$ and
$\bar {\cal Z} =   A^a_- A^a_-$. Consider the sector with a definite value of $j$. We can then write
 \be
\label{tildy}
P &=& \left( \frac {\cal Z}{\bar {\cal Z}} \right)^{j/4} \, \tilde P \, ,\nn
S &=& \left( \frac {\cal Z}{\bar {\cal Z}} \right)^{j/4} \, \tilde S \, , \nn
R &=& \left( \frac {\cal Z}{\bar {\cal Z}} \right)^{(j-2)/4} \, \tilde R \, ,\nn
T &=& \left( \frac {\cal Z}{\bar {\cal Z}} \right)^{(j-1)/4} \, \tilde T \, ,
 \ee
where $\tilde R, \tilde S, \tilde R$ and $\tilde T$ depend only on {\it two} neutral with respect
to the charge (\ref{j}) gauge-invariant
combinations ${\cal X}$ and ${\cal Y} = {\cal X}^2 - {\cal Z} \bar {\cal Z}$. In the vicinity of the valley,
they are reduced to ${\cal X} \to a^2$ and ${\cal Y} \to 4a^2 b^2 \ll {\cal X}^2$.

Let us act now on the wave function (\ref{Psibos}) by the supercharges (\ref{QQbar}). We obtain a system
of PDE of the first order for four functions $\tilde P, \tilde S, \tilde R, \tilde T$. One of these equations 
is actually algebraic, $\tilde T = 0$. Three remaining functions satisfy three equations.
 \be
\label{Qnatildy}
 {\cal X} \tilde S + \sqrt{{\cal X}^2 - {\cal Y}} \, \tilde R = i\lambda (\partial_{\cal X}  + 2{\cal X}
\partial_{\cal Y}) \tilde P + im \tilde P\, , \nn
\tilde S + \frac {\cal X}{\sqrt{{\cal X}^2 - {\cal Y}}} \, \tilde R = 2i\lambda \left[ \partial_{\cal Y} - \frac j{4({\cal X}^2 - {\cal Y})} 
\right] \tilde P \, , \nn
 \tilde P = -i\lambda \left\{ (\partial_{\cal X}  + 2{\cal X} \partial_{\cal Y})  \tilde S 
+ 2\sqrt{{\cal X}^2 - {\cal Y}}  \, \left[ \partial_{\cal Y} + \frac {j-2}{4({\cal X}^2 - {\cal Y})} 
\right] \tilde R \right \} + im \tilde S
\ee   
with $\lambda = 4g_0^2$. 
We are interested in the behavior of the solutions near the corner when ${\cal X} \ll \lambda/m$ [see (\ref{rangeC})]. In this region,
 mass terms are not important and can be neglected.
Let us do it and concentrate on the first two equations in (\ref{Qnatildy}) in the region ${\cal Y} \ll {\cal X}^2$ not
necessarily assuming that ${\cal X} \gg g_0^{2/3}$ (the condition for the Abelian BO description to be valid).
One can say that we are approaching the origin along the {\it trace} of the Abelian valley. Neglecting
${\cal Y}$ compared to ${\cal X}^2$, we readily see that the function $\tilde P({\cal X}, 0)$ satisfies the equation
 \be
\label{eqtildeP}
 \left( \frac  {\partial}{\partial {\cal X}} + \frac j{2{\cal X}} \right) \tilde P \ =\  0  
 \ee
with the solution
 \be
\label{soltildeP}
\tilde P \ \sim \ {\cal X}^{-j/2} \, .
 \ee
The behavior (\ref{soltildeP}) coincides with (\ref{Psi0a-j}) derived earlier. 
But it was derived there only in the region where the Abelian BO approximation is valid. 
The analysis of the full non-Abelian equations
(\ref{Qnatildy}) allowed us here to extend this asymptotics down to the very origin.

\section*{Appendix D. Supercharges, Hamiltonian and gauge-invariant variables.}
\setcounter{equation}0
\def\theequation{D.\arabic{equation}}

The Hamiltonian analysis of Sect. 3 can alternatively be done by  resolving the gauge constraints
 at the classical level and expressing the supercharges and Hamiltonian into gauge-invariant variables.
For the gauge SQM system obtained by reduction from (3+1) SYM theory, this was done in
\cite{howto}. We present here (mainly for methodical purposes) a similar analysis for the Hamiltonian
(\ref{Ham}). 

Six dynamical variables $A^{\ a}_j$ involve 3 gauge-invariant variables and 3 gauge angles. The latter
can be effectively separated if using the polar representation \cite{Simonov} ,
 \be
\label{polar}
A^{\ a}_j \ =\ U_{jk} \Lambda^{\ b}_k V_{ba} \, ,
 \ee
where $U_{jk} (\alpha)$ is an $O(2)$ matrix describing spatial rotations, $V_{ba} (\phi^a)$ is an $O(3)$ gauge
rotation matrix and $\Lambda^{\ b}_k$ is a quasidiagonal matrix,
\be
\label{Lambda}
\Lambda_k^{\ b} \ =\ \left( \begin{array}{ccc} a & 0 & 0 \\ 0 & b & 0 \end{array} \right)\ .
 \ee

By a  proper  spatial and/or gauge rotation the eigenvalues of $\Lambda^{\ b}_k$ can be brought
 to the range $a \geq |b|$. The fields (\ref{Lambda}) with positive or negative sign of $b$
are related to each other by a spatial reflection.  

Gauge-invariant variables are thus $a, b, \alpha$, while $\phi^a$ are gauge angles. 
The quantum problem involves, generally speaking, two sectors: the even in $b$ and odd in 
$b$ wave functions. In the leading BO approximation, the wave functions are even (see Eq.
(\ref{Psifastb}) below).

There are two ways to derive the expressions for gauge-invariant quantum supercharges 
and the Hamiltonian.
First, one can resolve the constraints at the classical level and obtain classical 
gauge-invariant 
supercharges. For supersymmetry to be kept at the quantum level, 
one should resolve the ordering ambiguities in the {\it supercharges} using symmetric 
Weyl prescription.
The quantum Hamiltonian is then obtained as the anticommutator $\{\bar Q, Q\}/2$. 
\footnote{Note that this Hamiltonian {\it does} 
not coincide with the Weyl-ordered classical Hamiltonian (see Ref.\cite{howto} 
for further details).}
Such supercharges and the Hamiltonian act in the Hilbert space with ``flat'' measure  
$\sim da db d\alpha $. However, the 3-manifold of gauge-invariant variables $(a,b,\alpha)$ is in fact
curved. If one is interested in the operators acting on the wave function
normalized with the covariant measure
 \be
\label{measureabalph}
  \prod_{aj} d A^a_j \ = a|b|  (a^2 - b^2) da db d\alpha d\mu_V 
 \longrightarrow \ C a|b|  (a^2 - b^2) da db d\alpha \equiv
\mu_{ab}\, da db d\alpha \ ,
 \ee 
($d\mu_V$ is the Haar measure on $SO(3)$), 
 one should perform a proper similarity transformation and replace $Q^{\rm flat}$ by
 \be
\label{similarity}
Q^{\rm cov} =  \frac 1{\sqrt{\mu_{ab}}} Q^{\rm flat} \sqrt{\mu_{ab}} 
 \ee
and, similarly, other operators.

Another approach is more direct and does not come to grips with a difficult ordering ambiguities problem,
 \begin{enumerate}
\item We take the expressions (\ref{QQbar}), (\ref{Ham}) for the quantum supercharges and Hamiltonian
and express them into new variables.
 \item Anticipating the eventual gauge fixing $\phi^a = 0$, we only consider a simplified version of 
these expressions in the small $\phi^a$ region such that the body-frame gauge angular momenta
 $J^a = V^{ad} \epsilon^{dbc} A^b_j E^c_j$, in terms of which the Hamiltonian (\ref{Ham}) is expressed,
go over to the generators $ J^a \to -i \partial /\partial \phi^a$. 
 \item The Gauss law constraints $\hat G^a \equiv 0$ allow one to express the latter  via the fermion variables, 
 \be
\label{Jviapsi}
J^a \ \equiv \ i \epsilon^{abc}  \psi^b \bar \psi^c \ .
 \ee
\end{enumerate}

This or other way, one obtains, assuming $b >0$,
\be
\label{Qresolved}
Q^{\rm cov} = e^{-i\alpha} g_0 \left[  \psi^1 \left( p_a - 
 i\frac {a p_\alpha +  b J^3 }{a^2 - b^2} - \frac {ima}{2g_0^2}\right) +
\psi^2 \left( -ip_b  + \frac {b p_\alpha +   
a J^3 }{a^2 - b^2} - \frac {mb}{2g_0^2}\right) \right. \nonumber \\
\left. - \psi^3 \left( \frac {J^2}a + \frac {i J^1}b \right)  \right] + 
\frac {iab}{g_0} \bar \psi^3 \ , \nonumber \\
\bar Q^{\rm cov} =  g_0 e^{i\alpha} \left[   \bar \psi^1 \left( 
p_a  + i\frac {a p_\alpha + 
 b J^3}{a^2 - b^2} +  \frac {ima}{2g_0^2} \right) +
 \bar \psi^2 \left( ip_b + \frac {b p_\alpha   + a J^3}{a^2 - b^2} -  
\frac {mb}{2g_0^2} \right)  \right. \nonumber \\
  -  \left.  \bar \psi^3  \left( \frac {J^2 }a - \frac {iJ^1}b \right)\right]
 - \frac {iab}{g_0} \psi^3 \ ,
 \ee
where $p_a = -i \partial/\partial a$, etc, and one should substitute for $J^a$ the fermion 
bilinears (\ref{Jviapsi}).
The Hamiltonian is
 \be
\label{Hresolved}
H \ =\ -\frac {g_0^2}2 \triangle -     \frac {im}2 \frac \partial {\partial \alpha} +
\frac 1{2g_0^2} \left[a^2b^2 + \frac {m^2}4 (a^2 + b^2)\right]
\nonumber \\
+ i\frac {\epsilon^{abc} }2 \left[ \bar \psi^a \bar \psi^b A^c_{+} +  \psi^a  \psi^b A^c_{-} \right]
 + \frac m2 \left( \psi^a \bar \psi^a - \bar\psi^a \psi^a \right) \ ,
 \ee
where 
$$A^1_\pm  = ae^{\pm i\alpha}\, , \ \ \ \ A^2_\pm  = \pm ibe^{\pm i\alpha} \, , \ \ \ \ \ A^3_\pm  = 0 \, .$$
 and 
  \be
 \label{Lapl}
 \triangle = \frac {\partial^2}{(\partial A_i^a)^2} \ =\ \frac {\partial^2}{(\partial a)^2} +
 \frac {\partial^2}{(\partial b)^2} + 
 \frac 1 a \frac {\partial} {\partial a} +  \frac 1 b \frac {\partial} {\partial b}
 + \frac 2{a^2 - b^2} \left( a \frac {\partial} {\partial a} - b \frac {\partial}  {\partial b} \right)
 \nonumber \\
+ \frac 1 {(a^2 - b^2)^2}
\left[ (a^2 + b^2) \left( \frac {\partial^2}{\partial \alpha^2} - (J^3)^2 \right) + 4iab \,   
J^3 \frac \partial {\partial \alpha}  \right] - \frac {(J^2)^2} {a^2} -  \frac {(J^1)^2} {b^2} \ ,
 \ee

The Hamiltonian (\ref{Hresolved}) is Hermitian with respect to the measure (\ref{measureabalph}),
$H^\dagger = \mu_{ab} H \mu_{ab}^{-1}$.   
The supercharges (\ref{Qresolved}) satisfy $Q^\dagger = \mu_{ab} \bar Q \mu_{ab}^{-1}$. They 
are nilpotent and their anticommutator 
gives (\ref{Hresolved}), as it should. These 
operators act on the wave functions normalized with the measure (\ref{measureabalph}).
The conserved angular momentum  (\ref{j}) is expressed as
 \be
\label{jresolved}
 j \ =\ p_\alpha + \frac 12 \psi^a \bar \psi^a \ .
 \ee
The expressions (\ref{Qresolved}), (\ref{Hresolved})  look complicated, 
but they are simplified a lot along
 the valleys. 
The slow bosonic variables are $a$ and $\alpha$. 
The combinations $a e^{\pm i\alpha}$ coincide with the variables $C_\pm$ of Sect. 3. Now,
$b^2$ is the fast variable, it corresponds to $(b^a_j)^2 $ of Sect. 3. 
The BO approximation works when $b_{\rm char}^2 \ll a^2$, which is true
as long as $a \gg g_0^{2/3}$ as in (\ref{rangeC}). 
The fast massless Hamiltonian (\ref{Hfast}) is expressed as
  \be
\label{Hfastb}
H^{\rm fast} =  -\frac {g_0^2}2\left[ \frac {\partial^2}{(\partial b)^2} +   
\frac 1 b \frac {\partial} {\partial b}\right] +
\frac
{  a^2 b^2}{2g_0^2} - \frac {g_0^2(\bar \psi^2 \psi^3 - \bar \psi^3 \psi^2)^2}{2b^2} + 
\nn 
\frac {ia}2 \epsilon^{ab}
\left[ e^{i\alpha} \bar \psi^a \bar \psi^b  +  e^{-i\alpha} \psi^a  \psi^b  \right] \, ,
 \ee
 where  $a,b = 2,3$ and $\epsilon^{23} = 1$. 
\footnote{Do not mix the color indices $a,b$  with the bosonic variables $a,b$  
(sorry, but the Latin alphabet is not large\dots).}

 The fast ground state wave function (\ref{Psifast}) takes the form
 \be
\label{Psifastb}
\Psi^{\rm fast} (b, \psi^2, \psi^3) \ \sim \ 
\frac 1 {\sqrt{a}} \exp \left\{- \frac {a b^2 }{2g_0^2}  \right\}
\left(  2i  +   e^{-i\alpha}  \epsilon^{ab}
  \psi^a \psi^b \right) \, .
 \ee
The analysis of Sect. 4 remains intact. The analysis of Appendix C can also be translated
into new variables, ${\cal X}, {\cal Y} \to a,b$.

\end{document}